# Strange metal behavior of the Hall angle in twisted bilayer graphene

Rui Lyu[1]*, Zachary Tuchfeld[1]*, Nishchhal Verma[1]*, Haidong Tian[1], Kenji Watanabe[2], Takashi Taniguchi[2], Chun Ning Lau[1], Mohit Randeria[1†], Marc Bockrath[1†]

[1]*Department of Physics, The Ohio State University, Columbus, OH 43210, USA*

[2]*National Institute for Materials Science, Namiki Tsukuba Ibaraki 305-0044 Japan*

*These authors contributed equally to this work.
†Corresponding author. Email: Bockrath.31@osu.edu (M.B.);
randeria.1@osu.edu (M.R.)

Abstract:

Twisted bilayer graphene (TBG) with interlayer twist angles near the magic angle ≈1.08º hosts flat bands and exhibits correlated states including Mott-like insulators, superconductivity and magnetism. Here we report combined temperature-dependent transport measurements of the longitudinal and Hall resistivities in close to magic-angle TBG. While the observed longitudinal resistivity follows linear temperature $T$ dependence consistent with previous reports, the Hall resistance shows an anomalous $T$ dependence with the cotangent of the Hall angle $\cot\Theta_H \propto T^2$. Boltzmann theory for quasiparticle transport predicts that both the resistivity and $\cot\Theta_H$ should have the same $T$ dependence, contradicting the observed behavior. This failure of quasiparticle-based theories is reminiscent of other correlated strange metals such as cuprates.

Transport in strongly correlated systems, where electronic quasiparticle excitations are not well-defined, has attracted attention in diverse areas of physics ranging from quantum materials (*1, 2*) and cold atoms (*3*) to string theory (*4*). Many of these investigations have their roots in the first observation of resistivity scaling linearly with temperature $T$ in the normal state of high temperature cuprate superconductors, along with a host of other anomalies including the $T^2$ scaling of cotangent of the Hall angle $\Theta_H$ (*5, 6*), which remain challenging open problems to this day.

Magic-angle twisted bilayer graphene (TBG) has emerged as a system exhibiting a wealth of tunable many-body phases (*7-17*). Moreover, the discovery of superconducting domes in the vicinity of correlated insulating phases (*8, 9*) shows that this system shares features with the high $T_c$ cuprate phase diagram (*9*). Nevertheless, there are many differences with the cuprates: the TBG electronic structure has multiple bands crossing the Fermi energy, Berry phase effects and Dirac dispersion. There are also differences in the phenomenology, such as the observation of nearby ferromagnetic phases. In view of this, it is important to understand the normal state transport phenomenology in TBG. Based on the linear $T$ resistivity, Cao et al. (*18*) have suggested that TBG exhibits exotic "Planckian dissipation" (*19*) like the cuprates, while other work (*20, 21*) has interpreted this behavior within a conventional electron-phonon scattering picture (*22*). In this Report, we argue that weak-field Hall resistivity and cotangent of the Hall angle add crucial insights to the story, and perhaps, can settle the debate.

Our main goal here is to address the question: can we understand *both* the observed longitudinal and Hall resistivities $\rho_{xx}$ and $\rho_{xy}$ within a conventional, textbook picture of metallic transport, or do the experiments imply an exotic transport regime characteristic of correlated quantum matter?

Here we study devices based on twisted bilayer graphene with interlayer twist angles near the magic twist angle $\approx 1.08°$. The longitudinal resistivity $\rho_{xx} \propto T$ over a range of temperatures from 10 K to 100 K. At higher temperatures we typically observe a resistivity maximum at a characteristic temperature $T_{max}$ ~100-200 K. The Hall resistance $\rho_{xy}$ is temperature dependent, and we observe that cot $\theta_H = \rho_{xx}/\rho_{xy} \propto T^2$ for the range over which the resistivity is *T*-linear.

While the linear-*T* $\rho_{xx}$ behavior can be consistent with electronic quasiparticle scattering by phonons, we find that this conventional picture cannot even qualitatively explain the Hall transport. Quite generally, in Boltzmann theory $\rho_{xx} \sim 1/\tau$, where $\tau$ is the transport scattering rate, while $\cot \theta_H = (\omega_c \tau)^{-1}$ where $\omega_c$ is the cyclotron frequency. Thus both $\rho_{xx}$ and $\cot \theta_H$ should exhibit identical *T*-dependence, arising from a common scattering rate (see Supplementary Material). The observed $\cot \theta_H \propto T^2$ behavior in the density and temperature regime where $\rho_{xx} \sim T$ is simply inconsistent with any quasiparticle-based transport theory. We note that this dichotomy is exactly the same as that found in the strange metal regime of high $T_c$ cuprates (*5, 6*).

Samples are made using the "tear and stack" technique on a custom-made micro-positioning stage similar to the method reported in ref. (*9*). A tBLG stack with a small twist angle ~1-2° is produced by tearing a graphene flake and rotating the separated pieces on top of each other. Figure 1A shows a schematic diagram of the layer stack showing the hexagonal BN-encapsulated device along with the oxidized Si wafer which acts as a gate electrode to modulate the charge density. The stack is then etched into a Hall bar geometry and Cr/Au edge contacts attached by electron beam lithography. A device optical image with attached electrodes is shown in the Fig. 1B inset. A top gate of $Al_2O_3$/Cr/Au is then evaporated on top of the final device in the region shown by the dashed line.

Completed samples are then loaded into a flowing He gas cryostat and transport measurements are performed. Figure 1B shows the longitudinal resistivity $\rho_{xx}$ vs. density *n* at the base temperature of 1.5 K. A peak is observed near charge neutrality, as well as for hole doping around a carrier density $n_0$ ~ -2.9 × $10^{12}$ cm$^{-2}$. This agrees with previous measurements, where because of the moiré superlattice produced by the interlayer twist, flat bands develop near charge neutrality, with an energy gap separating these bands from dispersive bands. When 4 electrons per moiré unit cell are removed the Fermi level reaches the hole side energy gap, producing a resistance peak. From the measured $n_0$, we infer an interlayer twist angle ~1.1°. (*8, 9*)

Figure 2A shows a color plot of the longitudinal resistivity $\rho_{xx}$ vs. *n* and *T,* while Fig. 2B shows $\rho_{xy}$ vs. *n* and *T*. Resistivity $\rho_{xx}$ vs. *T* extracted from Fig. 2A at different densities is shown by the blue black and red dashed lines in Fig. 2C. The corresponding colors show that above *T*~5 K $\rho_{xx}$ increases linearly with temperature. Data from another device D2 (similar color plots to Figs. 2A-B are shown in the Supplementary Material) are shown by the circles taken at several different densities. A similar linear *T* increase of $\rho_{xx}$ is evident. Figure 2D shows line traces of $\rho_{xy}$ for D1 and D2 taken at similar densities normalized to their corresponding full band filling. For both samples at sufficiently high *T*, $|\rho_{xy}|$ shows a decreasing trend with temperature.

After a linear *T* increase, $\rho_{xx}$ reaches a broad maximum before turning downwards. This behavior is shown in Fig. 3A, which shows maxima at a characteristic temperature which decreases as the doping increases towards $n/n_0$=4. We attribute this maximum as occurring when higher energy bands begin to contribute to transport. In Fig. 3B, we plot the band structure from ref. (*23*), which goes beyond the original continuum model (*23-25*) and includes structural relaxation that leads to a significant gap between the flat and dispersive bands. At the measured twist angle of 1.1° this shows a narrow band with bandwidth *W* ~ 40 meV. The right panel of Fig. 3B shows the corresponding density of states. The Fermi function is plotted for *T*=150 K at a density $n$=1.5 × $10^{12}$ cm$^{-2}$ as the blue dashed line in Fig. 3B. As the full width at half maximum of the Fermi function energy derivative is ≈3.5$k_B T$, we expect the excited band to begin being populated when 3.5$k_B T$ ~ *W*, yielding *T* ~ 130 K, in good agreement with the observed 150 K near half filling. The density dependence of the resistivity maximum can also be understood using this band structure (see Supplementary Material).

We now analyze the linear *T* variation of our resistivity in terms of various models. The measured resistivity $\rho_{xx}$ is seen to be clearly smaller than ($h/e^2$), and thus TBG is not obviously in a "bad metal" regime beyond the Mott-Ioffe-Regel limit with $k_F l$ ~ 1 (where $k_F$ is the Fermi wavevector and *l* the inferred electronic mean free path) where the quasiparticle picture necessarily breaks down. It thus makes sense to ask if we can understand TBG transport within a conventional Boltzmann formalism of weakly interacting quasiparticles.

The very narrow bandwidth inferred theoretically, and from our data, is consistent with previous reports(*7-13, 23-25*). This raises the possibility that *T*-variations of the density-of-states as determined by the compressibility $dn/d\mu$ (where *n* is the electron density and $\mu$ is the chemical potential), negligible in conventional metals, might be important for transport in magic-angle TBG. This impacts transport in narrow bands through the energy-derivative of the Fermi function and through the *T*-dependence of $\mu$, even if the quasiparticle scattering rate (from impurities) is *T*-independent. Such effects become significant when $k_B T$ is of order the bandwidth and, using realistic band structure parameters, we find that they are not relevant for the linear-T resistivity between 10 and 100 K (see Supplementary Material).

Next, let us consider electron-phonon scattering, which accounts for linear-T resistivity in many metals over a range of temperatures $T_{min} < T < T_{max}$ with $T_{max}$ determined by the Mott-Ioffe-Regel (MIR) criterion and $T_{min}$ a fraction of the Debye temperature. This mechanism is *not* responsible for the linear *T* resistivity in strongly correlated materials dubbed "strange metals" (*1, 2*). Such materials often violate the MIR criterion with no apparent $T_{max}$ for linear resistivity and $T_{min}$ is often too low to be consistent with phonons. TBG is different on both counts. First, the MIR criterion is never violated in TBG, as discussed above. Second, as emphasized in ref. (*22*), a very low $k_B T_{min} \sim \hbar\omega(Q)$ is obtained in low density materials (with small $k_F$), where $\omega(Q)$ is the phonon frequency at momentum transfer Q = 2$k_F$.

We can fit our $\rho_{xx}$ data treating electron-phonon scattering (*22*) within a Boltzmann approach. The scattering rate $1/\tau$ acquires a linear *T*-dependence from the phonon occupation for $k_B T > \hbar\omega(2k_F)$ and thus $\rho_{xx} \sim 1/\tau \sim T$. A quantitative fit to the slope can be obtained using (*22*) the result $\rho_{xx} = \pi F D^2 k_B T / g e^2 \hbar \rho_m v_F^2 v_a^2 + \rho_0$, where *D* is the deformation potential, $v_F$ and $v_a$ are the Fermi and acoustic phonon velocities, $\rho_m$ is the sheet mass density of graphene, *g* = 4 is the band degeneracy, $\rho_0$ is a constant representing the residual resistance from impurity scattering, and *F* is a dimensionless factor that depends on the twist angle (see Supplemental Information). These fits are shown in Fig. 2C as the dashed lines, and the parameter values are consistent with previous reports (*20, 22*) and with that expected for $v_F$ from the band structure of Fig 3B for both D1 and D2 ($v_F$ = 7×10$^4$ m/s for sample D1 and 1×10$^5$ m/s for D2).

We now turn to a discussion of the predictions of the electron-phonon model for the Hall response. Fig. 2D shows the measured $\rho_{xy}$(T) which is certainly not the T-independent response one expects in a simple metal. It is instructive to analyze our transport data in terms of the Hall angle $\cot\Theta_H = \rho_{xx}/\rho_{xy}$, which within Boltzmann theory has the simple expression $\cot\Theta_H = (\omega_c \tau)^{-1}$ (see Supplementary Material). Recall Matthiessen's rule: the total scattering rate is the impurity scattering rate, which determines the *T=0* intercept, plus the inelastic T-dependent scattering rate. Motivated by this, we subtract out the *T=0* value of the cotangent of the Hall angle and plot $\Delta\cot\Theta_H(T)$ in Fig. 4.

The inset of Fig. 4. shows the measured $\Delta\cot\Theta_H(T)$ versus *T*, which has a quadratic temperature dependence (dashed line), while the electron phonon theory leads to a linear *T* dependence (blue line). The quadratic variation of $\Delta\cot\Theta_H(T)$ in all our devices is further emphasized in Fig. 4 where we plot our data versus *T*$^2$.

We find that while electron-phonon scattering can indeed account for the linear *T* longitudinal resistivity, it fails to explain the observed quadratic behavior of $\cot\Theta_H$. We emphasize that this problem is not limited to electron-phonon scattering *per se*, but points to a deeper failure of Boltzmann theory of quasiparticle transport. Within this theory the same scattering rate enters both the longitudinal and Hall conductivities (see Supplementary Material), so that $\rho_{xx}$ and $\cot\Theta_H$ must have the same $1/\tau$ variation and thus the same *T*-dependence. This is clearly at odds with our measurements.

We are forced to conclude that the totality of transport data in magic-angle TBG cannot be understood in terms of conventional quasiparticle transport theory, and appears to be very similar to – the still unsolved – problem of normal state transport in the high $T_c$ cuprate superconductors and other strongly correlated systems. This provides further support for the notion that correlations remain important in the normal metal phase above $T_c$ and may provide some insight into the mechanism of the formation of the correlated phases.

Our results raise the question of why quasiparticle transport theory breaks down in TBG. The absence of a scale – other than the temperature $k_B T$ -- in the scattering rate $\hbar/\tau$ could be understood in terms of a nearby quantum critical point (QCP) where other energy scales collapse to zero. QCPs are a common feature of many of the correlated materials where similar transport anomalies have been seen. QCPs are well established in heavy fermion materials (*26*) and Fe-based superconductors (*27*). In the cuprates too there is mounting evidence for a QCP (*28, 29*), whose nature is less well understood. It is thus an important open question whether there is an underlying QCP in magic angle-TBG that controls the anomalous power laws seen in our transport data.

**Acknowledgements**
**Funding:** The experimental portion of the work was funded by the Department of Energy under award DE-SC0020187. N.V. and M.R. were supported by the Center for Emergent Materials, an NSF-funded MRSEC under Grant Nos. DMR-1420451 and DMR-2011876. **Author Contributions** R. L. and Z. T. performed the experiments, and H. T. assisted with device fabrication. M. B. and C. N. L. supervised the experiments. N. V. and M. R. performed the theoretical analysis. K. W. and T. T. grew the BN material. All authors contributed to the writing of the manuscript.
**Competing Interests:** The authors declare no competing interests. **Data and materials availability:** Data shown in main text and Supplementary Material available upon reasonable request.

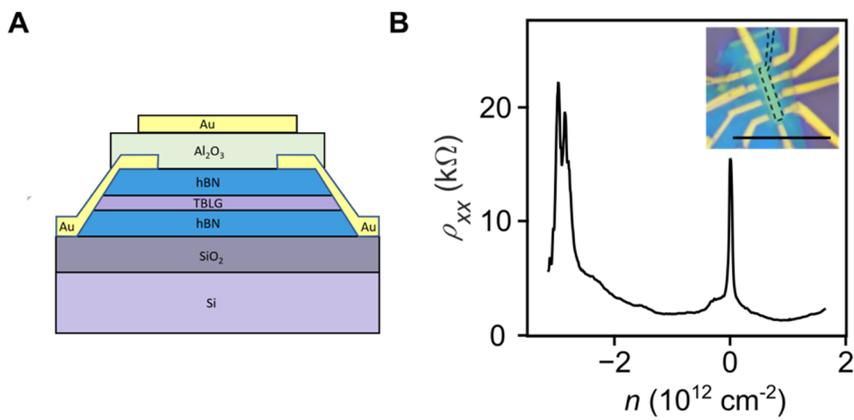

**Fig. 1. TBG device geometry and electronic transport measurements.** (**A**) Schematic of layer stack geometry showing a BN-encapsulated TBLG with attached gate and contact electrodes. (**B**) Plot of $\rho_{xx}$ vs. $n$ for a sample D1 with twist angle ≈1.1° taken at $T$=1.5 K. Inset: optical image of a different device D2 with twist angle ≈0.91°. Dashed line shows top gate location fabricated in a later step. Scale bar 20 μm.

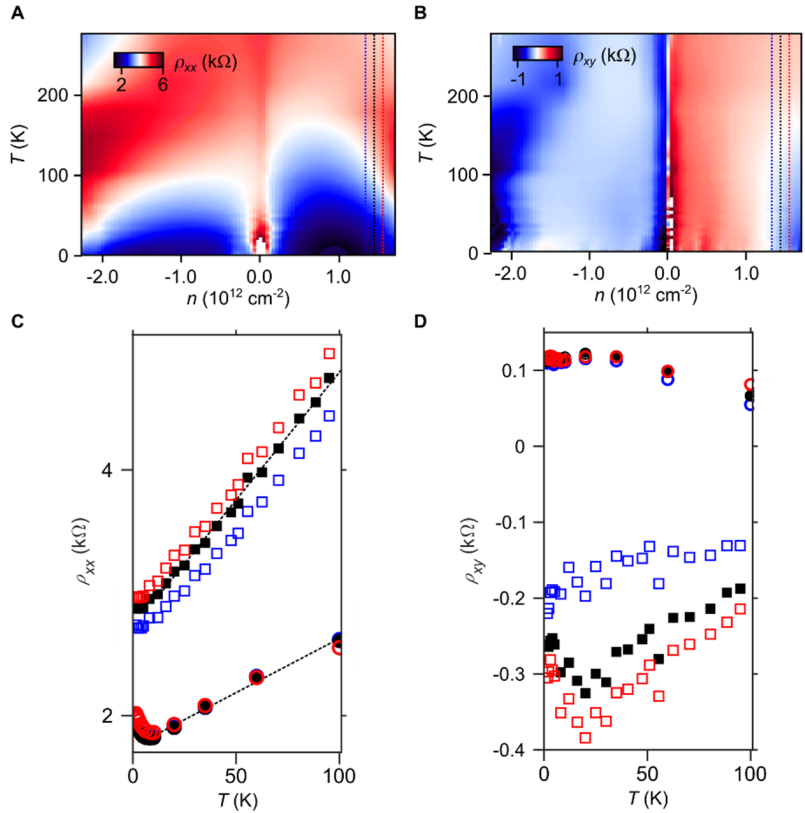

**Fig. 2. Longitudinal and Hall resistivity of TBG devices.** (**A**) Color plot of resistivity $\rho_{xx}$ vs. temperature $T$ and density $n$ taken from sample D1. (**B**) Color plot of resistivity $\rho_{xy}$ vs. $T$ and $n$, taken at a magnetic field $B = 0.2$ T. An offset has been subtracted from each density trace to make the low-density regime fit the classical Hall behavior, see supplement. (**C**) Line traces of $\rho_{xx}$ for two samples. The squares indicate data taken D1 with the line traces having the corresponding colors shown in part **A**. For clarity, this data has been offset upwards by 1 kΩ. Filled symbols indicate data taken near half-filling, while open symbols indicate data taken away from half-filling. The dashed line shows a linear fit to the data at half-filling, yielding a slope of 18 Ω/K. Circles indicate data from another device D2. The dashed line indicates a fit with a slope 8.8 Ω/K. (**D**) transverse resistivity $\rho_{xy}$ for the same two samples with data for D2 taken at $B=0.3$ T. Data was obtained using a contact symmetrization procedure, see for example ref. 13.

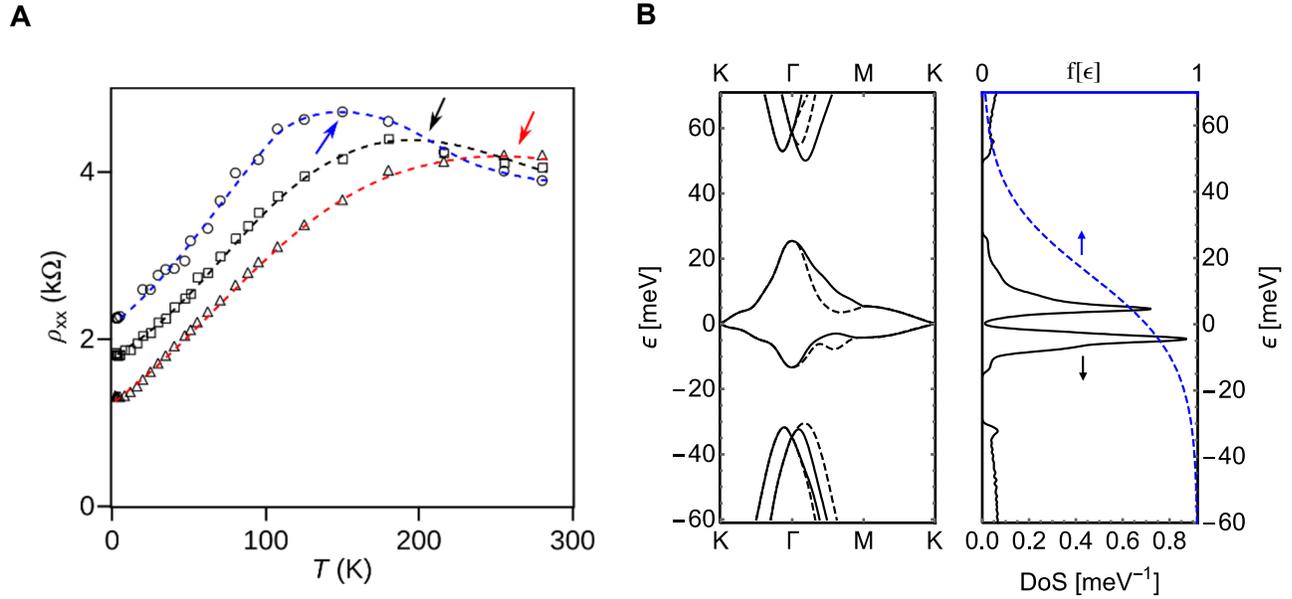

**Fig. 3. Temperature dependence of resistivity and theoretical model.** (**A**) $\rho_{xx}$ versus $T$ up to higher temperatures for D1, taken at densities $n$=0.96, 1.4, 1.64 × $10^{12}$ cm$^{-2}$ (triangles, squares, circles respectively). Dashed lines are guides to the eye. Each data trace shows a maximum at a temperature $T_{max}$ shown by the arrow with the corresponding color to each guide that decreases with density. (**B**) Calculated band structure for a twist angle 1.1° following ref. 23. The band structure comes with an estimate of the energy gap between the narrow and higher bands. The peak in resistivity can be understood as the activation of carriers in higher bands. Blue dashed line shows the Fermi function centered at chemical potential corresponding to density 1.5 × $10^{12}$ cm$^{-2}$, with $T$ = 150 K.

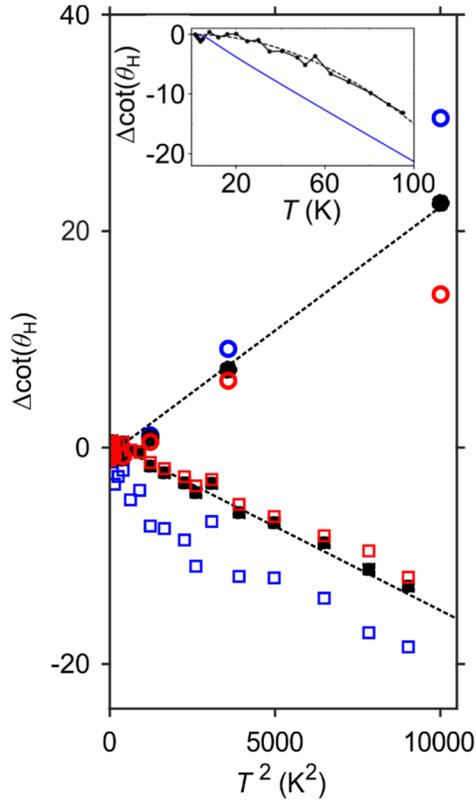

**Fig. 4. Temperature dependence of the cotangent of the Hall angle.** Main panel: Change in the cotangent of the Hall angle from its low temperature value versus $T^2$. Squares represent data from D1, while circles represent data from D2. Colors for D1 correspond to those in Fig. 2. Data for D2 is taken at similar band filling factors. Filled symbols indicate data taken near half-filling, while open symbols indicate data taken away from half-filling. The dashed lines show linear fits to the half-filling data. Inset: measured change in cot $\theta_H$ vs. $T$ near half-filling for D1 on a linear scale. Blue line shows the trend expected from a theory calculation using phonon scattering.

# Supplementary Material: Strange metal behavior of the Hall angle in twisted bilayer graphene

## (I) Additional data figure from sample D2 and analysis of $\rho_{xy}$ data.

**Sample D2 data**

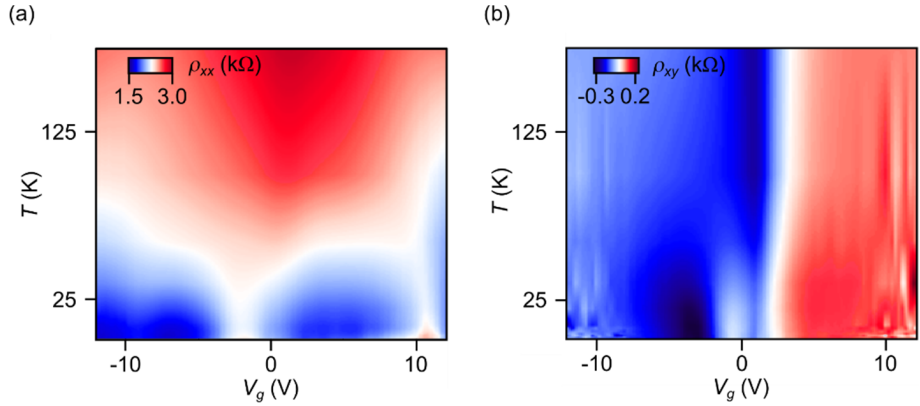

Fig. 1. (a) Color plot of resistivity $\rho_{xx}$ vs. temperature $T$ and gate voltage $V_g$ taken from sample D2. (b) Color plot of resistivity $\rho_{xy}$ vs. $T$ and $V_g$, taken at a magnetic field $B$ = 0.3 T.

**Analysis of $\rho_{xy}$ data**

For sample D1, an apparent offset in $\rho_{xy}$ was corrected for each gate voltage sweep by averaging the data over a range of charge densities $n$ near charge neutrality and subtracting this constant from each sweep. The resulting data follows the expected $\rho_{xy}=B/ne$ behavior close to charge neutrality. For sample D2, $\rho_{xy}$ was obtained using Onsager symmetrization as in, for example, ref. 1.

## (II) Can we use Boltzmann theory to understand the data?

Boltzmann transport theory of quasi-particle transport is valid in the regime $k_F l \gg 1$, where $k_F$ is the Fermi momentum and $l$ the mean-free path. Strange metals often – but not always – violate the Mott-Ioffe-Regel limit $k_F l \sim 1$. If that were the case, one cannot justify using Boltzmann theory to understand the data.

A natural question to ask is if this is the case with TBG. To answer that, recall that in 2D, $k_F l$ can be estimated directly from observed resistivity as $\rho_{xx} \sim h/e^2 \, (k_F l)$, where $h/e^2 = 25.81 \, k\Omega$ is the Von-Klitzing constant. We find that even at the highest temperatures (100K) up to which linear resistivity is observed in our TBG samples, the estimates of $k_F l$ range from 5 – 15.

It is thus *a priori* reasonable to use Boltzmann theory and ask if it can explain the data. We note, however, that at the end of our analysis we will conclude that a Boltzmann theory based on quasiparticle transport is *not* able to account for the observed temperature dependence of both $\rho_{xx}(T)$ and $\rho_{xy}(T)$, at least using any known quasiparticle scattering mechanisms.

# (III) Conductivities from linearized Boltzmann equation

We review the standard derivations for the longitudinal and Hall conductivities using the linearized Boltzmann equation within the relaxation time approximation. The distribution function $f_{mk}$ for electrons in band $m$ with crystal momentum $k$ is obtained by solving the Boltzmann equation

$$\frac{e}{\hbar}(E + v_{mk} \times B) \cdot \nabla_k f_{mk} = -\frac{(f_{mk} - f^0_{mk})}{\tau_{mk}} \tag{1}$$

Here $E = E\,\hat{x}$ and $B = B\,(-\hat{z})$ are the external fields, $v_{mk} = \hbar^{-1}\nabla_k \epsilon_{mk}$ is the band velocity, $f^0_{mk} = f^0(\epsilon_{mk})$ is the Fermi function and $\tau_{mk}$ is a phenomenological scattering time. Writing $f_{mk} = f^0_{mk} + g_{mk}$ and keeping terms to linear order in $E$ we find

$$\left[1 + \frac{e}{\hbar}\tau_{mk}(v_{mk} \times B) \cdot \nabla_k\right] g_{mk} = e\,\tau_{mk}(E \cdot v_{mk})\left(-\frac{\partial f^0_{mk}}{\partial \epsilon_{mk}}\right). \tag{2}$$

We solve for $g_{mk}$ by inverting the operator on the left-hand side above and keeping terms up to first order in B (*weak-field* limit). We thus find

$$g_{mk} = \left(1 - \frac{e}{\hbar}\tau_{mk}(v_{mk} \times B) \cdot \nabla_k + \cdots\right)\left(e\,\tau_{mk}(E \cdot v_{mk})\left(-\frac{\partial f^0_{mk}}{\partial \epsilon_{mk}}\right)\right) \tag{3}$$

from which obtain the current

$$J = e \sum_{mk} v_{mk}\, g_{mk}. \tag{4}$$

We then calculate the conductivity

$$\begin{pmatrix} J_x \\ J_y \end{pmatrix} = \begin{pmatrix} \sigma_{xx} & \sigma_{xy} \\ -\sigma_{xy} & \sigma_{yy} \end{pmatrix} \begin{pmatrix} E \\ 0 \end{pmatrix} \tag{5}$$

to find

$$\sigma_{xx} = \frac{e^2}{\hbar} \sum_{mk}\left(-\frac{\partial f^0_{mk}}{\partial \epsilon_{mk}}\right)(v^x_{mk})^2 \tau_{mk} \tag{6}$$

and

$$\sigma_{xy} = e^2\left(\frac{eB}{\hbar}\right)\sum_{mk}\left(-\frac{\partial f^0_{mk}}{\partial \epsilon_{mk}}\right)(v^y_{mk}\tau_{mk})\left(v^y_{mk}\frac{\partial}{\partial k_x} - v^x_{mk}\frac{\partial}{\partial k_y}\right)(v^x_{mk}\tau_{mk}). \tag{7}$$

Since experiments measure resistivities, we invert the conductivity matrix to find

$$\rho_{xx}(B=0) = \frac{1}{\sigma_{xx}}, \qquad \rho_{xy} = -\frac{\sigma_{xy}}{\sigma_{xx}^2 + \sigma_{xy}^2} \approx -\frac{\sigma_{xy}}{\sigma_{xx}^2}. \qquad (8)$$

TBG has four bands (2 spin x 2 valley) crossing the chemical potential at any given filling of the narrow bands. We simply take this into account using degeneracy factor of 4 in the equations above.

# (IV) Band structure

We need a model band structure to calculate the resistivities and there are many available in the literature (2–6). The continuum model of Bistritzer and McDonald correctly predicted the nearly flat bands at the magic angle but ignores lattice relaxation effects that lead to a large gap between the flat bands and higher bands. We use the band structure of Carr et al. (5) which includes the effects of lattice relaxation and uses maximally localized Wannier functions that respect the symmetries of TBG and the fragile topology of its electronic structure. Additionally, this model can be used for arbitrary twist angles, even away the magic-angle.

We show how the band structure of TBG permits us to understand two distinct qualitative aspects of the transport data: (i) non-monotonic T-dependence of $\rho_{xx}$, and (ii) variation of the low temperature $\rho_{xy}$ with density.

(i) Non-monotonicity of $\rho_{xx}(T)$: As already noted above, a realistic energy gap between the narrow bands and higher bands is built into the band structure of Carr et al.; see Fig. 2a. The Fermi occupancy factor in eq. (6) broadens with increasing temperature, and eventually carriers in the higher band begin to contribute to transport as shown in Fig. 3b of main text. This leads to and eventual decrease in the resistivity at high T, with a peak near 150K (for the parameters shown). This corresponds to half filling on the electron doped side of charge neutrality, using the bare band structure of Carr et al for the angle 1.12° relevant to our Device D1. In this simple picture, the temperature at which the resistivity peak occurs should increase upon doping away from half-filling towards charge neutrality. This is indeed observed in the data shown in Fig. 3a of main text.

(ii) Doping dependence of low temperature $\rho_{xy}$: We will see below that the Boltzmann approach is *not* able to account for the T-dependence of $\rho_{xy}$, however, it is able to say something useful about the doping dependence when the temperature T << bandwidth. In this limit, the **k**-sums in eqs. (6) and (7) are dominated by states near the Fermi surface (FS). Next, we convert the sum over **k** to an integral over energy together with a density of states $N(\epsilon)$. If we further assume $\tau_{m\mathbf{k}} = \tau(\epsilon_{m\mathbf{k}})$, we get the standard results (7)

$$\sigma_{xx} = 2\,e^2\,N(0)\,v_F^2\,\tau \qquad \text{and} \qquad \sigma_{xy} = (\omega_c \tau)\,\sigma_{xx} \qquad (9)$$

where $N(0)$ is the density of states at the Fermi level, $v_F$ is the velocity averaged over the FS, given by $v_F^2 = \langle v_k^2 \rangle_{k \in FS}$, and $\tau$ is the relaxation time at Fermi energy. The spin and valley degeneracy of 4 times ½ (in 2D) accounts for the numerical pre-factor in $\rho_{xx}$. In the Hall conductivity, we use the cyclotron frequency $\omega_c = eB/m^*$ with the effective mass $m^*$ given by the "average curvature" of the FS

$$\frac{1}{m^*} = \frac{\langle \sum_{ij} v_k^i [Tr(M^{-1}(k))\delta_{ij} - M_{ij}^{-1}(k)] v_k^j \rangle_{k \in FS}}{\langle \sum_{ij} v_k^i [\delta_{ij}] v_k^j \rangle_{k \in FS}} \quad (10)$$

where $M_{ij}^{-1}(k) = \hbar^{-2} \partial_i \partial_j \epsilon(k)$. For $\epsilon \sim |k|^n$ in 2D, we can derive the relation $m^* v_F = \hbar k_F$, which is particularly useful for Dirac (n=1) and parabolic (n=2) dispersions.

It is clear from these results that the sign of Hall resistivity in eq. (8) is determined by that of $m^*$, which change sign at a Lifshitz transition where the system goes from a particle-like to hole-like Fermi surface. The zero crossing of $\rho_{xy}$ can thus reveal important details about the band structure. We show in Fig. 2d that our choice of band structure captures this feature of the data.

## (V) Impurity scattering in narrow bands

Elastic scattering from impurities scattering is usually invoked to account for the temperature independent resistivity at low T. However, one might ask if impurity scattering in a very narrow band system can by itself lead to non-trivial T dependence in transport. While we will see that this is true in principle, it cannot explain any of the observed anomalies.

The elastic scattering rate off impurities is given by the Fermi golden rule expression

$$\Gamma_{mk} = \frac{\hbar}{\tau_{mk}} = 2\pi \sum_{m'k'} |M_{mk,m'k'}|^2 \delta(\epsilon_{mk} - \epsilon_{m'k'}) \quad (11)$$

where, $M_{mk,m'k'}$ is impurity matrix element for scattering from electron state $(m, k)$ to $(m', k')$. For a structureless matrix element, this simplifies to

$$\frac{1}{\tau_{mk}} = \frac{2\pi}{\hbar} n_{imp} U^2 N(\epsilon_{mk}) \quad (12)$$

where $n_{imp}$ is the density of impurities, $U$ is the strength of impurity potential and $N(\epsilon_{mk})$ is the density of states (DOS). The scattering rate is thus T independent but depends on DOS which is determined by the band structure. Using this result in eq. (6), we see that there are two factors that contribute to temperature dependence: (i) the T dependence of chemical potential µ(T) for a given density (see Fig. 2a), and (ii) the derivative of Fermi function which has a height $\sim 1/T$ and width $\sim T$.

At temperatures large compared to µ and the bandwidth, the sum in eq. (6) extends to all states in the band, and the $1/T$ pre-factor from $\left(-\frac{\partial f^0_{mk}}{\partial \epsilon_{mk}}\right)$ dominates the conductivity. This leads to a linear in $T$ resistivity. However, this linear T resistivity does *not* persist down to low temperatures. At temperatures small compared to µ and the bandwidth, we can approximate $\left(-\frac{\partial f^0_{mk}}{\partial \epsilon_{mk}}\right) \approx \delta(\epsilon_{mk} - \mu)$ in eq. (6), and then µ(T) controls the T-dependence of the resistivity. As µ increases with T, so does the Fermi velocity and this leads to a decrease in resistivity with increasing T as shown in Fig. 2b. However, we note that the variation of Fermi velocity with chemical potential depends on the details of the band structure, and is hence a non-universal feature. (In conventional metals, where bandwidth $W \sim 10^4 K$, the effects we are discussing are utterly negligible).

The upshot of this analysis is that these two effects together can lead to a non-trivial T dependence in a very narrow band even from a T-independent impurity scattering rate, but not one that can explain the observed data in TBG. In Fig. 2b, we show an (unsuccessful) attempt to describe the linear T data over some range of temperatures using such an approach. The strength of scattering is fixed by matching the *T = 0* resistivity. We see that renormalizing the Carr et al band dispersion $\epsilon_{mk} \to r\epsilon_{mk}$ by a factor of r = 0.4 we can "fit" the slope of linear T regime between 40K and 100K, but then we also get a low-T upturn below 40K from the µ(T) effect described above. Band renormalization can be used to tune the temperature of this crossover, but the overall non-monotonicity is robust.

The temperature dependence of $\rho_{xy}(T)$ further supports this conclusion argument. Narrow band effects, arising from the derivative of Fermi function, lead to $|\rho_{xy}| \sim T$, as shown in Fig 2c. The experimental data clearly has a different temperature dependence.

Therefore, despite the subtleties of very narrow band systems, impurities alone fail to explain the observed data in TBG. In fact, the observed monotonic behavior of $\rho_{xx}(T)$ implies that the T-dependence arising from Fermi factors is not relevant for TBG, and one must look at the T-dependence of the scattering rates in eqs. (6) and (7) to understand the data.

## (VI) Electron-Phonon scattering: Longitudinal Resistivity

Next, we turn to acoustic phonon scattering, which accounts for linear-T resistivity in metals over a range of temperatures $T_{min} < T < T_{max}$. Here $T_{min}$ is usually a fraction of the Debye temperature $\Theta_D$, but can be much smaller in low density systems (see below), while $T_{max}$ is essentially determined by the Mott-Ioffe-Regel criterion $k_F l(T_{max}) \sim 1$.

We emphasize that electron-phonon scattering is clearly *not* responsible for the linear T resistivity in strongly correlated materials that are dubbed "strange metals". In these systems the Mott-IR criterion is violated, with no observed $T_{max}$, and at the same time $T_{min}$ is too low to be explained using phonons. TBG is different on both counts. First, as discussed in Section I, the Mott-IR criterion is never violated in TBG, so there is no problem with $T_{max}$. In fact at high T other bands come into play, as discussed above in Section III. Further, as emphasized by Wu et al[8], $T_{min}$ is not as issue for $\rho_{xx}$ in TBG either; see below. We agree with these authors that one can understand $\rho_{xx}(T)$ in terms of electron-phonon scattering, however, we will show that this mechanism cannot explain the observed Hall response.

The onset $T_{min}$ of the linear-T scattering rate is controlled by the Bloch-Grüneisen temperature $k_B T_{BG} = \hbar v_{ph} |\Delta \boldsymbol{k}|_{max} = \hbar v_{ph} 2 k_F$, where $2 k_F$ is the maximum momentum transfer across the Fermi surface; see Fig. 3a. In conventional metals (with $k_F \sim \pi/a$) $T_{BG}$ is a fraction of the Debye temperature $\Theta_D$. However, in a low-density system, we see that $T_{BG}$ can be parametrically smaller than $\Theta_D$. For $T \gg T_{BG}$, the phonon occupation $n(\omega) \sim T/T_{BG}$ and this in turn leads to an electron-phonon scattering rate $\Gamma \sim T$.

This phenomena has been studied in detail in Dirac materials like graphene[9] where for small densities and $T \gg T_{BG}$

$$\rho_{xx} = \frac{\pi}{e^2 \hbar v_F^2} \left[ \frac{1}{\rho_m} \left( \frac{D}{v_{ph}} \right)^2 k_B T \right]. \tag{13}$$

Here $D$ is the deformation potential (electron-phonon coupling), $\rho_m$ the mass density and $v_F$ the Dirac velocity. (Here the scattering rate is the quantity in square brackets). The model has been extended to TBG by including the renormalization of the Dirac velocity and changes in electron-phonon matrix elements. As shown in eq. (2) of Wu et al. (8), the resistivity is then given by

$$\rho_{xx} = \frac{4}{e^2 (v_F^*)^2} \left[ F(\theta) \frac{1}{\rho_m} \left( \frac{D}{v_{ph}} \right)^2 (v_{ph} k_F) \, I\left(\frac{T}{T_{BG}}\right) \right]. \tag{14}$$

Here the dimensionless functions $F(\theta)$, coming from the matrix element of electron-phonon Hamiltonian, and $I(z)$ given in terms of a definite integral, are the same as those defined in ref.[8]; their numerical values are plotted in Fig 3a. We followed the procedure outlined in their appendix A to calculate $F(\theta)$ for $\theta = 1.12°$. All other factors were taken to be the same as that of monolayer graphene, keeping $v_F^*$ as the only fit parameter. The best fit to our experimental data was obtained with $v_F^*/v_F \approx 0.057$ where $v_F = 10^6 \, m/s$ is the Dirac velocity of monolayer graphene. This is very close to the prediction of the continuum model ($v_F^*/v_F \sim 0.05$).

As emphasized by Wu et al. (8), electron-phonon scattering is the simplest mechanism that explains both resistivity slope *($\sim 100 \, \Omega/K$)* and the early onset of linear-T resistivity ($\sim 5 \, K$) within a conventional Boltzmann formalism. We show next that the same model fails to provide even a qualitative explanation of the T-dependence of the Hall response.

## (VII) Electron-Phonon scattering: Hall response

We next show that phonons, and indeed Boltzmann transport, simply fail to explain the temperature dependence of cot $\Theta_H$. We see from eq. (9) that

$$cot\ \Theta_H = (\omega_c \tau)^{-1} \qquad (15)$$

is a general consequence of Boltzmann transport, since the same scattering mechanism impacts both $\sigma_{xx}$ and $\sigma_{xy}$. Thus $\rho_{xx}$ and $cot\ \Theta_H$ share the same $1/\tau$ dependence, and hence the same temperature dependence.

We note that in the paper we plot

$$\Delta cot\ \Theta_H = cot\ \Theta_H(T) - cot\ \Theta_H(0) \qquad (16)$$

motivated by Matthiessen's rule: $1/\tau$ is the sum of the impurity scattering rate $1/\tau(T=0)$ plus the inelastic scattering rate. Thus $\Delta cot\ \Theta_H$ allows us to focus on the latter and ignore the *T=0* impurity contribution.

*Any* inelastic scattering mechanism (including electron-phonon scattering) that explains linear-T resistivity within a quasiparticle picture has a rate that can be written as

$$\Gamma_{m,\mathbf{k}} = \frac{\hbar}{\tau_{m,\mathbf{k}}} = k_B T\ \Phi_{m,\mathbf{k}} \qquad (17)$$

where $\Phi_{m\mathbf{k}}$ depends on the band structure and scattering matrix elements. but *not* on T. When the *k*-sums in the Boltzmann results are dominated by FS contributions, only the value of $\Phi_{m\mathbf{k}}$ evaluated at the FS affects the pre-factors, but the temperature dependence of $cot\ \Theta_H$ is simply linear in T. This is clearly in qualitative disagreement with the experimentally observed $cot\ \Theta_H \sim T^2$.

We emphasize that observed behavior cannot be attributed to *multi-band* effects in the Boltzmann theory of Hall response (see, e.g., Sec. 12.2 of Ref. [6]). The resistivity clearly requires that the different bands have the same linear T-dependence in the scattering rates, even if the numerical coefficients were to be band-dependent (see eq. (17)). Then, even though the final multi-band result for the Hall resistivity is complicated, the T-dependence simply cancels out.

In conclusion, while acoustic phonon scattering can account for the "strange" linear T resistivity, it fails to explain the quadratic behavior of $cot\ \Theta_H$. The failure is not limited to phonons but is equally problematical for any scattering mechanism within Boltzmann theory. Therefore, we argue that TBG is indeed a "strange" metal, whose "strangeness" is perhaps not directly visible in the longitudinal resistivity but is emphasized by considering the cotangent of Hall angle together with $\rho_{xx}$.

# Figures

Fig 2:

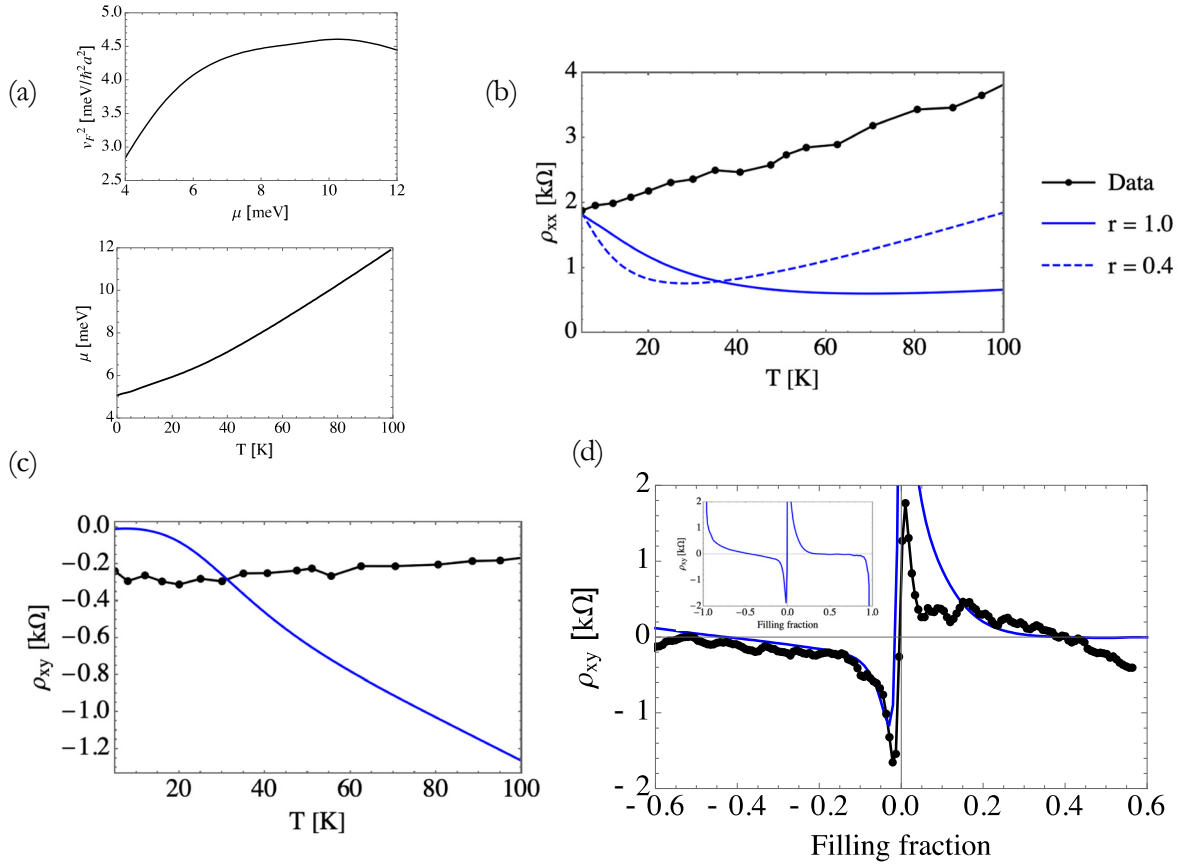

(a) Trend of Fermi velocity squared vs chemical potential. In the range $4-10\ meV$, $v_F$ increases with $\mu$. $\mu(T)$ itself increases with temperature. This results in a bigger Fermi velocity which increases conductivity and decreases resistivity. The phenomenon is more pronounced for smaller bandwidths.
(b) Numerically calculated Resistivity vs Temperature. Model calculations (blue) show non-monotonicity which is not observed in the data (black). A renormalization factor can change the minima but cannot get rid of non-monotonicity. Strength of scattering rate is $20\ meV$ and $0.45\ meV$ for $r = 1.0$ and $0.4$ respectively.
(c) Temperature dependence for Hall resistivity which mainly comes from the derivative of the Fermi function. However, the dependence is wrong.
(d) Density dependence of Hall resistivity at low temperatures. Inset: Density dependence for all filling fractions of the narrow bands.

Fig 3:

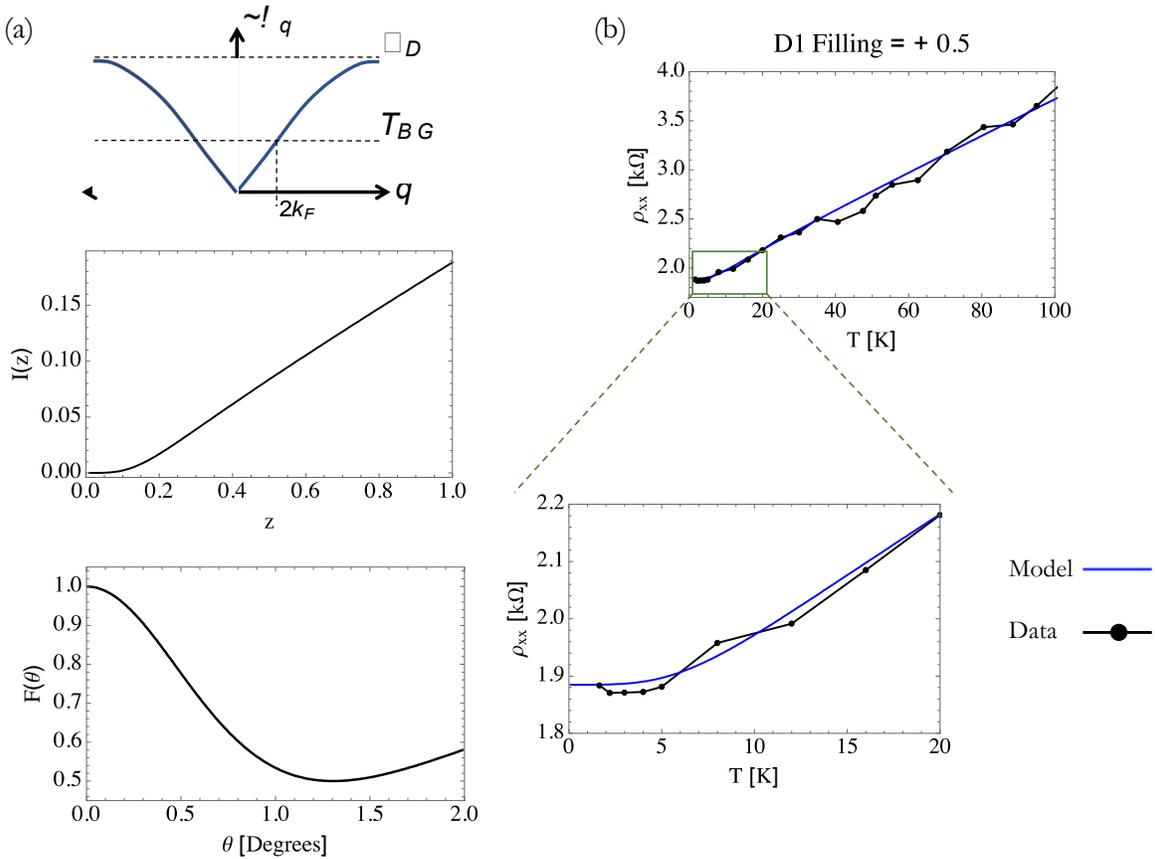

(a) Schematic phonon dispersion showing the difference between Debye temperature and Bloch-Gruneisen temperature for a system with small density. Band structure factor $F(\theta)$ and Bloch-Gruneisen integral $I(z)$ in eq. (14). $I(z)$ is linear for $z \gg 1$ and the linearity sets in at $z \approx 0.25$. This gives $T_{min} = T_{BG}/4$. $F(\theta)$ is a number bounded between 0.5 and 1.
(b) Phonon model fit (blue) to data (black) for device D1. The model captures both, slope and early onset of linearity with $v_F^*/v_F = 0.057$.